# Impact of Hydrostatic Pressure on Superconductivity of $Sr_{0.5}La_{0.5}FBiS_2$

Rajveer Jha, Brajesh Tiwari, and V.P.S. Awana[*]

CSIR-National Physical Laboratory, Dr. K.S. Krishnan Marg, New Delhi-110012, India

We report the impact of hydrostatic pressure (0–1.97Gpa) on superconductivity of recently discovered 2K superconductor $Sr_{0.5}La_{0.5}FBiS_2$. Resistivity under pressure measurements are performed by using HPC-33 Piston type pressure cell with Quantum design DC resistivity Option. The superconducting transition temperature ($T_c$) is increased by 5 fold to around 10K with just above 1GPa pressure, which remains nearly unaltered for studied higher pressures of up to 1.97GPa. The fivefold increase in $T_c$ accompanied with decrease in normal state resistivity of $Sr_{0.5}La_{0.5}FBiS_2$ with just above 1GPa pressure calls for the attention of solid state physics community.

[*]**Corresponding Author**
Dr. V. P. S. Awana, Principal Scientist
E-mail: awana@mail.nplindia.org
Ph. +91-11-45609357, Fax-+91-11-45609310
Homepage www.fteewebs.com/vpsawana/

Discovery of superconductivity in $BiS_2$ layers of $Bi_4O_4S_3$ [1,2] and $ReO/FBiS_2$ [3-10] has attracted tremendous interest of both experimentalists and theoreticians from condensed matter physics community. The principal reason for it is the structural similarity of $ReO/FBiS_2$ with the famous $ReO/FFeAs$ pnictides high $T_c$ superconductors [3-12]. Though the $T_c$ of $BiS_2$ superconductors is say below 10K only, but their structural similarity with layered low dimensional and strongly electron correlated high $T_c$ cuprates (HTSc) and pnictides warrants their in-depth study. One of the issues argued by the theoreticians is that though in HTSc cuprates and pnictides the role of 3d metal (Cu/Fe) is more important, in case of $BiS_2$ based systems the same role is played by the sp orbitals of Bi and S [13,14]. Comparing the role of various Re in $ReO/FFeAs$ [11-15] and $ReO/FBiS_2$ [3-8], one finds that $T_c$ increases monotonically from La, to Pr and Nd in both the systems. The interesting comparison though is that $T_c$ seems to be scaled by factor of ten from FeAs to $BiS_2$ superconductors. For example the $T_c$ of (La/Pr/Nd)O$_{0.5}$F$_{0.5}$BiS$_2$ is around 2.1K/3.5K/5K respectively [3,4,5], the same for (La/Pr/Nd)O$_{0.8}$F$_{0.2}$FeAs is 26K/43K/50K respectively [11,13,14]. To an extent it appears that chemical pressure plays an important role in superconductivity of these layered BiS$_2$ and FeAs superconductors. In fact the same is true for HTSc Cuprates as well, where the replacement of La by Y in the Nobel Prize winning LaBaCuO$_x$ compound [16], could enhance the $T_c$ from around 30K to 90K [17] for YBa$_2$Cu$_3$O$_7$.

It is clear that traditionally the internal chemical pressure by on site substitutions or the externally applied pressures had played a major role in discovery of new superconductors. As far as the BiS$_2$ based newest superconductors are concerned, couple of detailed studies are available related to impact of hydrostatic pressure on their superconductivity [18-21]. In particular, though the $T_c$ of Bi$_4$O$_4$S$_3$ decreases [18, 19] with external pressure, it increases tremendously for the ReO/FBiS$_2$ [18-21]. In BiS$_2$ based superconductor family, the parent compound is recently been proposed to be un-doped SrFBiS$_2$ [22]. The substitution of Sr$^{2+}$ by La$^{3+}$ dopes mobile carriers in the BiS$_2$ layer and superconductivity is achieved with $T_c$ of above 2K [23-25]. Ironically, the impact of external pressure on the superconductivity of Sr$_{0.5}$La$_{0.5}$FBiS$_2$ is not yet studied. In this short communication, we report the impact of external pressure on the superconductivity of Sr$_{0.5}$La$_{0.5}$FBiS$_2$ superconductor. The superconducting transition temperature ($T_c$) of around 2K for Sr$_{0.5}$La$_{0.5}$FBiS$_2$ superconductor is enhanced to above 10K for just above 1GPa external pressure. The results are new and thought provoking for condensed matter scientific community.

The bulk polycrystalline Sr$_{0.5}$La$_{0.5}$FBiS$_2$ sample was synthesized by solid state reaction route via vacuum encapsulation. High purity La, SrF$_2$, Bi and S were weighed in stoichiometric ratio and ground in pure Argon atmosphere glove box. The mixed powder was subsequently palletized and vacuum-sealed (10$^{-4}$ mbar) in a quartz tube. The box furnace was used to sinter the sample at 650$^0$C for 12h with the typical heating rate of 2$^o$C/min. The sintered sample was subsequently cooled down slowly to room temperature. This process was repeated twice. X-ray diffraction (*XRD*) was performed at room temperature in the scattering angular (*2θ*) range of 10$^o$-80$^o$ in equal *2θ* step of 0.02$^o$ using *Rigaku Diffractometer* with *Cu K$_α$* ($λ$ = 1.54Å). Rietveld analysis was performed using the standard *FullProf* program.

The pressure dependent resistivity measurements were performed on Physical Property Measurements System (*PPMS*-14T, *Quantum Design*) using HPC-33 Piston type pressure cell with Quantum design DC resistivity Option. Hydrostatic pressures were generated by a BeCu/NiCrAl clamped piston-cylinder cell. The sample was immersed in a fluid pressure transmitting medium of Fluorinert in a Teflon cell. Annealed Pt wires were affixed to gold-sputtered contact surfaces on each sample with silver epoxy in a standard four-wire configuration. The pressure at low temperature was calibrated from the superconducting transition temperature of Pb.

The room temperature observed and Reitveld fitted XRD patterns of as synthesized $Sr_{0.5}La_{0.5}FBiS_2$ sample are shown in Figure 1. The sample is fitted in tetragonal structure with centrosymmetric space group *P4/nmm*. The refined lattice parameters are $a=4.077(2)$Å and $c=13.388(3)$Å. XRD results are in good agreement with previous reports on $Sr_{0.5}La_{0.5}FBiS_2$ compound [22, 23]. Strontium/Lanthanum (Sr/La), Bi and Sulfur (S1 and S2) atoms occupy the 2*c* (0.25, 0.25, *z*) site with $z=0.110(2)$ for Sr/La, 0.622(3) for Bi, 0.375(2) for S1 and 0.813(2) for S2. On the other hand F atoms are located at 2*a* (0.75, 0.25, 0) site. The atomic coordinates, Wyckoff positions, and site occupancy for studied $Sr_{0.5}La_{0.5}FBiS_2$ sample are given in Table.1. The resultant unit cell of the $Sr_{0.5}La_{0.5}FBiS_2$ compound with *P4/nmm* space group is shown in the inset of the Fig 1. Various atoms with their respective positions are indicated in crystal structure. The layered structure is composed of a fluorite-type SrF and stacked rocksalt-type $BiS_2$ layer along the *c* axis.

The temperature dependence of the electrical resistivity (ρ) below 250K down to 2K for $Sr_{0.5}La_{0.5}FBiS_2$ compound at various applied pressures of 0.35GPa-1.97GPa along with the data without applied pressure are shown in Figure 2a. The normal state resistivity behavior of the compound without applied pressure is clearly semiconducting down to superconductivity onset of around 2.2K, and superconductivity is seen with $T_c(\rho=0)$ at 2K. The normal state semiconducting behavior along with superconductivity at 2K is in agreement with reported data on $Sr_{0.5}La_{0.5}FBiS_2$ superconductor [22, 23]. With applied pressure of 0.35GPa the normal state behavior of the resistivity changes to metallic and the same improves further for higher applied pressures till 1.38GPa. For higher pressures of above 1.38GPa the normal state behavior is though metallic, but with slightly less metallic slope. As far as normal state resistivity ($\rho^{20K}$), i.e., well above superconducting transition temperature onset is concerned the same is 10mΩ-cm for zero pressure, 0.128mΩ-cm for 0.35GPa, 0.106mΩ-cm for 1.38GPa, and 0.153mΩ-cm for 1.68GPa and 1.98GPa pressures. Clearly the normal state resistivity at 20K, when compared with 0GPa sample, is decreased by more than an order of magnitude with 0.35GPa pressure, remains within same range for up to 1.38GPa and later increases slightly for 1.68GPa and 1.97GPa pressures. The decrease in normal state resistivity ($\rho^{20K}$), by more than an order of magnitude with 0.35GPa pressure along with the change of conduction process from semiconducting to metallic is surprising. It is possible that under pressure F-Sr/La-F bond angle along with inter-atomic distances do change, which in turn affect the charge density at Fermi surface and hence a clear change in normal state electrical transport.

As far as superconductivity i.e., $T_c(\rho=0)$ is concerned the same can be seen clearly from Fig. 2(b), which is zoomed part of Fig. 2(a) near superconducting transition state. The $T_c(\rho=0)$ for $Sr_{0.5}La_{0.5}FBiS_2$ sample without pressure is around 2K, which with applied pressure of 0.35GPa, and 0.55GPa is nearly the same but increases sharply to 8.6K for 0.97GPa. With further increase in pressure to 1.38GPa, 1.68GPa and 1.97GPa, the $T_c(\rho=0)$ is enhanced to above 10K. It is clear that superconducting transition temperature $T_c(\rho=0)$ remains nearly unchanged at around 2K for applied pressure till 0.55GPa and later increases sharply to around 10K for further higher pressure of 0.97GPa and remains nearly same (10K) till 1.97GPa. The plot of superconducting transition temperature [$T_c(\rho=0)$] with the applied pressure is shown in inset of Fig. 2(b). The fivefold increase in $T_c$, with applied pressure of just above 1GPa is surprising and calls for the attention of solid state physics community. Fivefold increase in $T_c$ and large decrease in normal state resistivity at small pressure (~1GPa) suggests possible structural transformation and change in band structure in $Sr_{0.5}La_{0.5}FBiS_2$ as similar to $LaO_{0.5}F_{0.5}BiS_2$ compound [26].

Summarizing, the results of Fig. 2(a) and 2(b), It has been observed that with increasing pressure the normal state resistance turns to the metallic behavior instead of the semiconducting along with a fivefold increase in superconducting transition temperature from 2K to 10K. This is though qualitatively same, but slightly different quantitatively from earlier reports on pressure dependent superconductivity of other $BiS_2$ based superconducting $(La/Pr/Nd/Ce)O_{0.5}F_{0.5}BiS_2$ compounds [18-21]. In particular, though nearly fivefold increase was seen for superconducting transition temperature [20], the normal state conduction along with normal state resistivity is not improved so dramatically as in present case of $Sr_{0.5}La_{0.5}FBiS_2$ superconductor. Worth mentioning is the fact that this is first study on impact of hydrostatic pressure on superconductivity of $Sr_{0.5}La_{0.5}FBiS_2$ superconductor. Interestingly, $SrFBiS_2$ is proposed to be the parent compound for the $BiS_2$ based superconductor family [22]. The coexistence of accompanied insulator to metal transition and fivefold increase in $T_c$ of $Sr_{0.5}La_{0.5}FBiS_2$ under moderate pressure of just above 1GPa warrants further detailed studies related to structural transformation under pressure in this newest class [1-10, 18-24] of $BiS_2$ based superconductors.

Figure 2(c) shows the temperature dependent resistivity of $Sr_{0.5}La_{0.5}FBiS_2$ superconductor at 0GPa and 1.97GPa pressure from 250K to 1.9K. The normal state resistance of $Sr_{0.5}La_{0.5}FBiS_2$ at 0GPa exhibits clearly the semiconducting behavior down to superconducting onset at 2.3K. At zero pressure the Resistivity versus temperature clearly shows the thermally activated behavior from 250K down to 2.3K. The thermal activation energy ($\Delta$) is obtained by fitting with thermal activation equation $\rho(T)= \rho_0 \exp(\Delta/2k_BT)$ for the temperature range 200K to 90K and is shown in the inset of Fig.2(c). The activation energy is estimated about 10.2 meV, which suggests a small energy gap in this compound. Similar semiconducting behavior of the resistivity has earlier been reported for $LaO_{0.5}F_{0.5}BiS_2$ [3]. Interestingly, the charge density wave instability has been suggested recently [27] by the first-principles calculations, which may cause enhanced electron correlations in this system. The normal state resistivity shows metallic behavior as applied pressure increases. Here it can be clearly seen that the highest pressure 1.97GPa sample exhibit metallic conductivity with positive $d\rho/dT$ along with a superconducting transition at below 10K. Red solid line shows the linearly fitted resistivity plot according to equation $\rho=\rho_0 + AT$, where $\rho_0$ is the residual resistivity and A is the slope of the graph. The normal state resistance data for pressure 1.97GPa is well fitted in linear equation in the temperature range 240K to 20K. The values of $\rho_0$ and A are found to be $1.34\times10^{-4}\Omega$-cm and $15.7\mu\Omega/K$ respectively. This behavior is quite similar to the iron based superconducting compound $NdFeAsO_{0.8}F_{0.2}$ [28] and different from some other superconductors like $MgB_2$, and some of the iron based compounds. In these compounds the resistivity $\rho$ shows quadratic temperature dependence, $\rho = \rho(0) + AT^2$, in the temperature range $70K \leq T \leq 150K$, and a linearly dependence in T intermediate region, $170K \leq T \leq 270K$ for $PrFeAsO_{0.6}F_{0.12}$ [29].

Figure 3 shows the temperature dependent resistivity under applied magnetic fields (1-7Tesla) for 1.97GPa pressure $Sr_{0.5}La_{0.5}FBiS_2$ sample. Both the superconducting transition $T_c$ onset and $T_c(\rho=0)$ decrease with increasing applied magnetic field. It is observed that $T_c^{onset}$ decreases less compared to $T_c(\rho=0)$ with the increasing magnetic field. The transition width is broadened in this compound. The upper critical field $H_{c2}$ versus T for of 1.97GPa pressure for the $Sr_{0.5}La_{0.5}FBiS_2$ sample is shown in Figure 4. The upper critical field $H_{c2}$ is estimated by using the conventional one-band Werthamer–Helfand–Hohenberg (*WHH*) equation, i.e., $H_{c2}(0) = -0.693T_c(dH_{c2}/dT)_{T=Tc}$. The $H_{c2}$ corresponds to the temperatures, where the resistivity drops to 90%, 50% and 10% of the normal state resistivity $\rho_n(T,H)$ at $T_c^{onset}$ in applied magnetic fields. The solid lines are the result of fitting of $H_{c2}(T)$ to the *WHH* formula. The estimated $H_{c2}(0)$ are 20 Tesla for 90%, 11.5 Tesla for 50%, and 8.5 Tesla for 10% $R_n$ criteria.

In conclusion, we have synthesized phase pure tetragonal $Sr_{0.5}La_{0.5}FBiS_2$ superconductor with superconductivity at 2K and studied the impact of hydrostatic pressure on its superconductivity and normal state conduction. It is found that superconducting transition temperature of 2K is increased fivefold to above 10K under just above 1GPa pressure and is accompanied with a normal state semiconductor to metallic transformation as well. Suitable chemical pressure by on site substitutions in recently discovered $BiS_2$ superconductors may enhance their superconductivity to high $T_c$ regime. This is because their dTc/dP of ~ 8K/GPa is unique and surprising as well.

Authors would like to thank their Director NPL India for his keen interest in the present work. This work is financially supported by *DAE-SRC* outstanding investigator award scheme on search for new superconductors. Rajveer Jha acknowledges the *CSIR* for the senior research fellowship.

**Table 1**
Atomic coordinates, Wyckoff positions, and site occupancy for studied $Sr_{0.5}La_{0.5}FBiS_2$.

| Atom | x | y | z | site | Occupancy |
|---|---|---|---|---|---|
| Sr/La | 0.2500 | 0.2500 | 0.110(2) | *2c* | 0.5/0.5 |
| Bi | 0.2500 | 0.2500 | 0.622(3) | *2c* | 1 |
| S1 | 0.2500 | 0.2500 | 0.375(1) | *2c* | 1 |
| S2 | 0.2500 | 0.2500 | 0.813(2) | *2c* | 1 |
| F | 0.7500 | 0.2500 | 0.000 | *2a* | 1 |

**Figure Captions**

**Figure 1:** (Color online) Observed (*open circles*) and calculated (*solid lines*) XRD patterns of $Sr_{0.5}La_{0.5}FBiS_2$ compound at room temperature.

**Figure 2(a):** (Color online) Resistivity versus temperature ρ(T) plots for $Sr_{0.5}La_{0.5}FBiS_2$ compound, at various applied pressures in the temperature range 250K-2.0K.

**Figure 2(b):** (Color online) Resistivity versus temperature ρ(T) plots for $Sr_{0.5}La_{0.5}FBiS_2$ compound, at various applied pressures in the temperature range 20K-2.0K, inset shows the $T_c$ offset versus pressure plots for $Sr_{0.5}La_{0.5}FBiS_2$ compound.

**Figure 2(c):** (Color online) Resistivity versus temperature ρ(T) plots for $Sr_{0.5}La_{0.5}FBiS_2$ compound, at 0GPa and 1.97GPa pressure in the temperature range 250K-2.0K. Inset shows the thermal activated fitted normal state resistance in the range 200K to 90K.

**Figure 3:** (Color online) Temperature dependence of the Resistivity ρ(T) under magnetic fields for the $Sr_{0.5}La_{0.5}FBiS_2$ compound under 1.97GPa hydrostatic pressure.

**Figure 4:** (Color online) The upper critical field $H_{c2}$ taken from 90%, 50% and 10% Resistivity criterion ρ(T) for the $Sr_{0.5}La_{0.5}FBiS_2$ under 1.97GPa hydrostatic pressure.

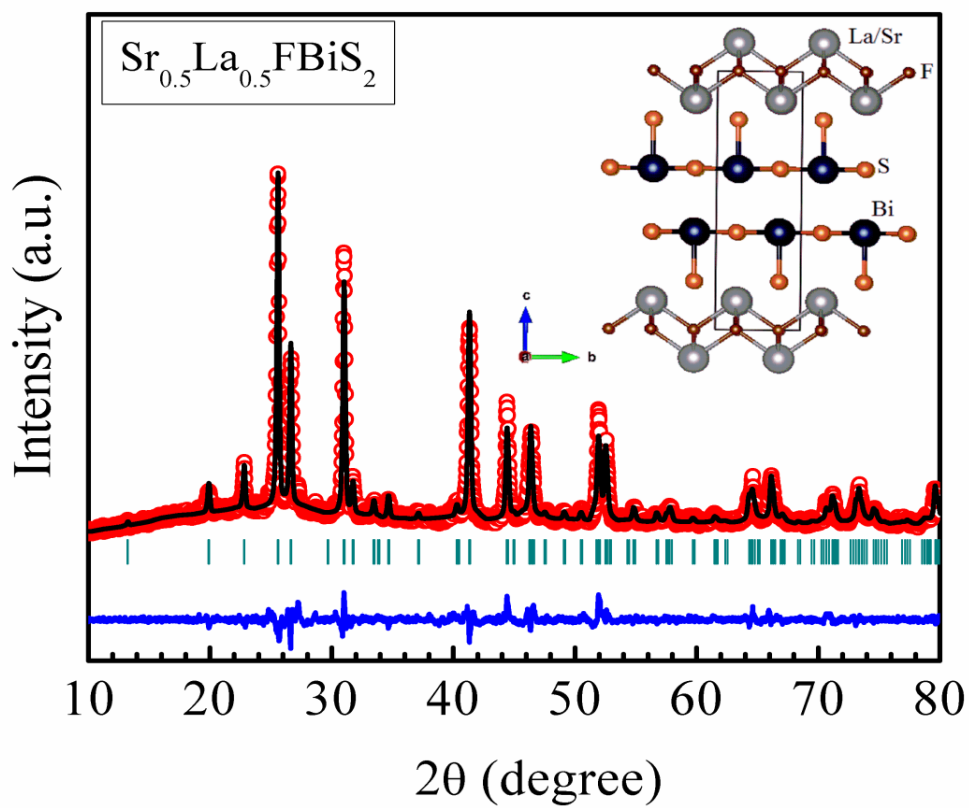

Figure 1

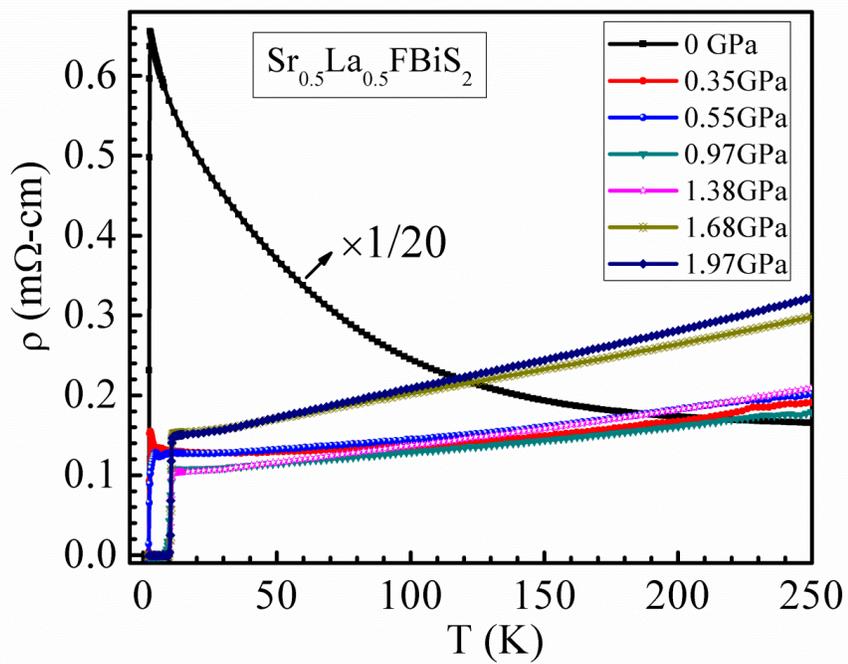

Figure 2(a)

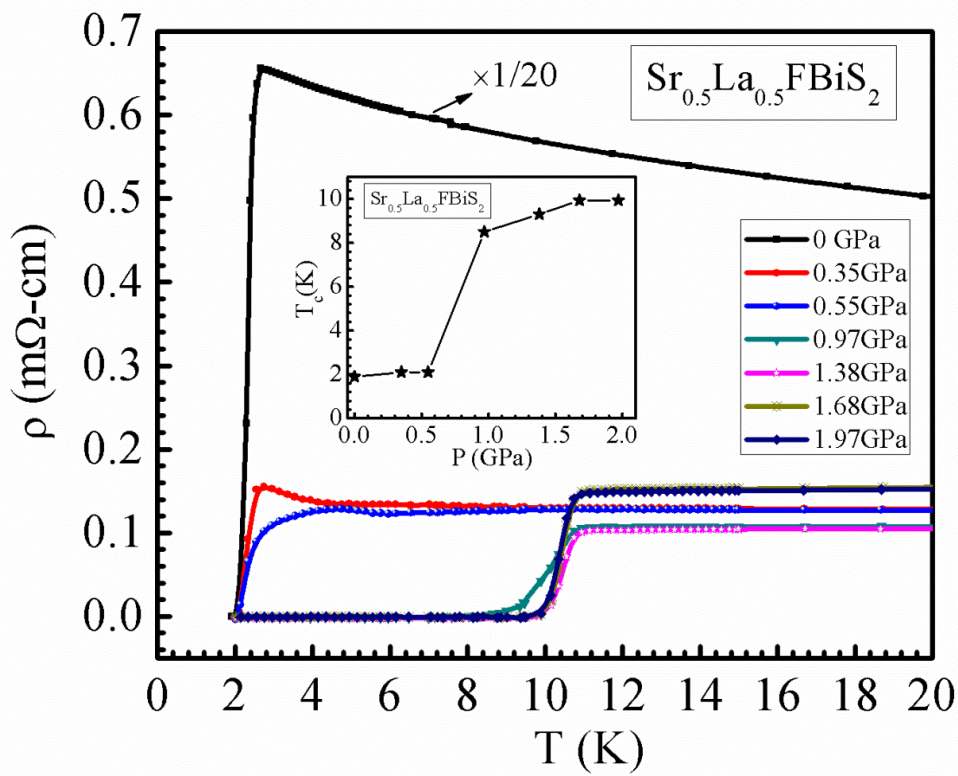

Figure 2(b)

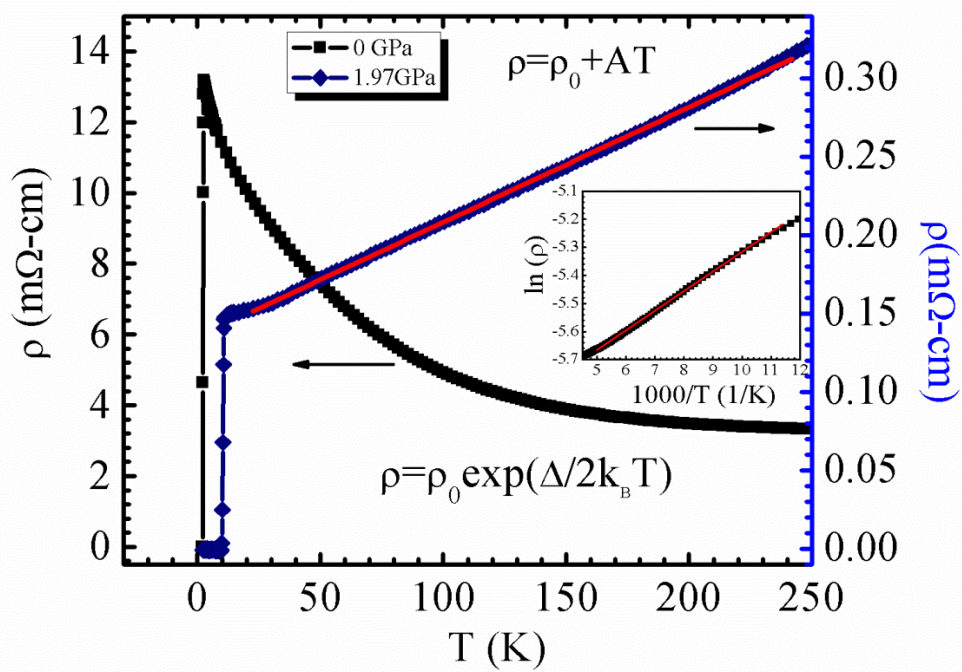

Figure 2(c)

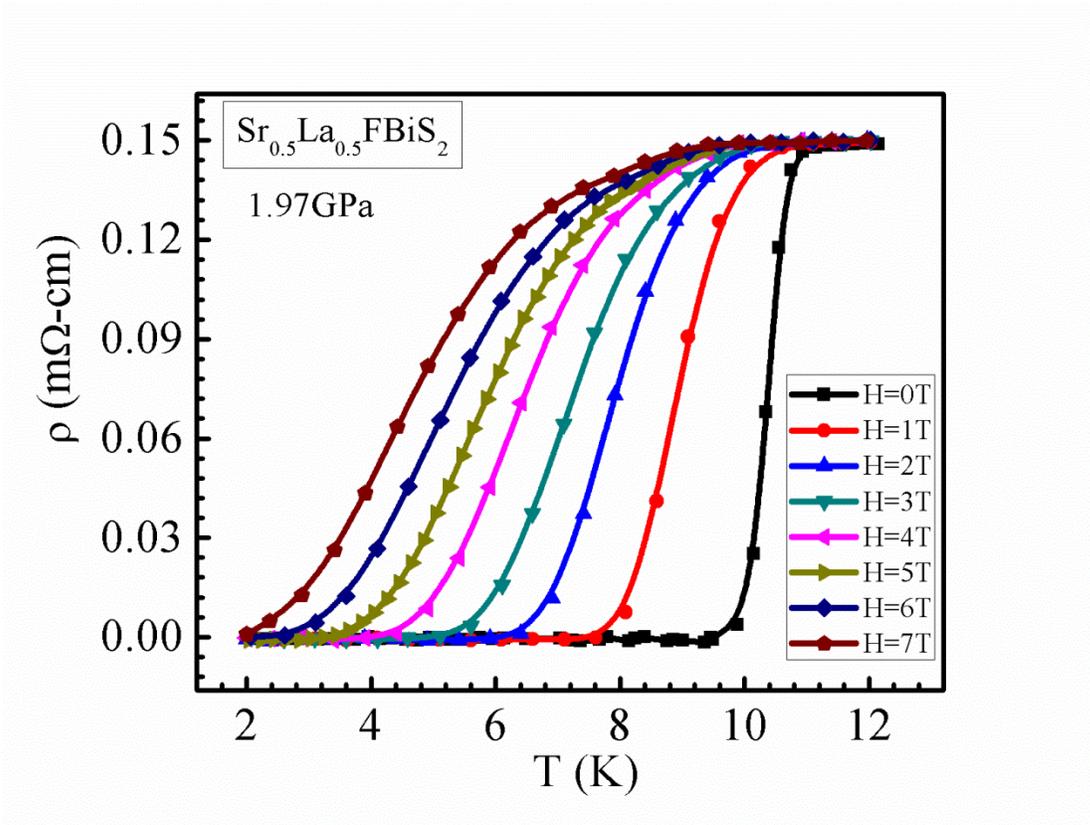

Figure 3

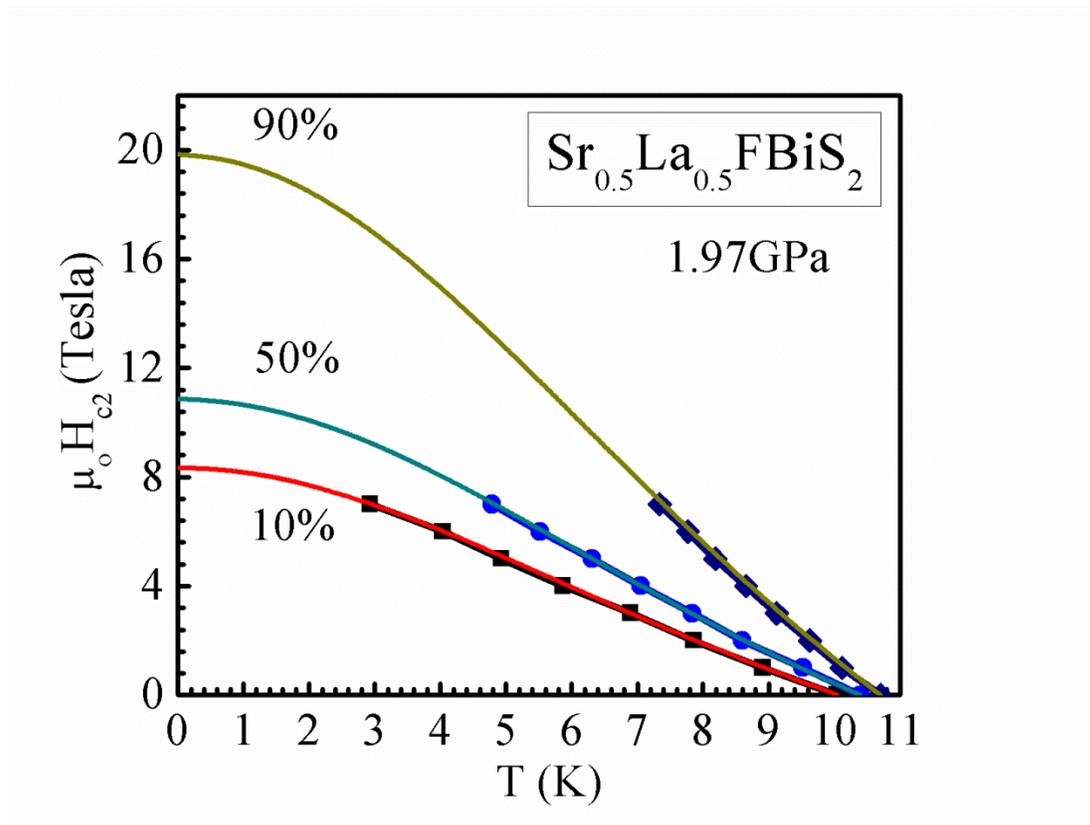

Figure 4